\documentclass[12pt,a4paper]{article}
\usepackage[english]{babel}
\usepackage{graphicx}

\newcommand{\alt}{\mathbin{\lower 3pt\hbox
   {$\rlap{\raise 5pt\hbox{$\char'074$}}\mathchar"7218$}}}
\newcommand{\agt}{\mathbin{\lower 3pt\hbox
   {$\rlap{\raise 5pt\hbox{$\char'076$}}\mathchar"7218$}}}

\begin{document}
\setcounter{footnote}{0}
\setcounter{equation}{0}
\setcounter{figure}{0}
\setcounter{table}{0}

\begin{center}
{\large\bf Critical Exponents from Field Theory: \\
New Evaluation. }

\vspace{4mm}
A. A. Pogorelov, I. M. Suslov \\
P.L.Kapitza Institute for Physical Problems,
\\ 119337 Moscow, Russia \\
 E-mail: suslov@kapitza.ras.ru
\vspace{1mm}
\end{center}

\begin{center}
\begin{minipage}{135mm}
{\large\bf Abstract } \\
We present   new evaluation of the critical exponents of
$O(n)$--symmetric $\phi^4$ theory  from the field
theoretical renormalization group, based on the new algorithm for
summing divergent series. The central values practically coincide
with those by Le Guillou and Zinn-Justin (1980) but their
uncertainty is essentially smaller.

 \end{minipage} \end{center}

\vspace*{1.5mm}
\vspace*{1.5mm}

\vspace{6mm}
\begin{center}
{\bf 1. Introduction}
\end{center}

The present paper has an aim to give  new evaluation
of the critical exponents of $O(n)$--symmetric $\phi^4$ theory
from the field theoretical renormalization
group (RG) \cite{1}, based on the new algorithm for summing
divergent series \cite{0}.

According to the formalism of the field-theoretical RG,
one should calculate three functions $\beta(g)$, $\eta(g)$,
$\eta_2(g)$ entering the Callan-Symanzik equation, find a
non-trivial root  $g^*$ of equation $\beta(g)=0$ (determining the
fixed point of
the RG equations), and then the critical exponents
$\eta$ and $\nu$, as well as the exponent $\omega$ of correction
to scaling, are given by expressions
$$
\eta=\eta(g^*)\,,\qquad \nu^{-1}=2-\eta(g^*)+\eta_2(g^*)
\,,\qquad  \omega=\beta'(g^*)\,. \eqno(1)
$$
The RG functions are given by factorially divergent series
in powers of the coupling constant $g$ and to calculate them one
need a method for summing divergent series. The examples are
Pade-Borel \cite{2} or conformal Borel \cite{3} techniques.

Our initial information is the same as in the paper
\cite{4}, i.e. the first 7 expansion coefficients of the RG
functions $\beta(g)$, $\eta(g)$, $\eta_2(g)$  \cite{2,4} and
their large order behavior \cite{5} established
in the framework of the Lipatov method \cite{6}.
The main difference from the preceding papers consists in the
fact that  explicit interpolation of the coefficient
function is made from the very beginning: the low order
expansion coefficients are smoothly interpolated with their large
order asymptotics and unknown intermediate coefficients are found
in a certain approximation. Considering these coefficients as
exact, one can sum the divergent series with (in principle)
arbitrary precision. The only uncertainty of the algorithm is
related with ambiguity of interpolation, which has a clear
physical sense and originates from incompleteness of initial
information.
%
%
As a consequence,
 the relation of the summation results with
 the assumed behavior of the coefficient functions can
 be constructively analyzed and estimation of uncertainty
 becomes completely transparent.
 For technical reasons, such procedure was impossible
 in the conventional algorithms due to catastrophic
 increase of errors in the course of the series resummation,
 which made interpolation to be useless. A crucial
 point is  stability of our algorithm with respect
 to smooth errors, involving  ambiguity of
 interpolation.

\vspace{6mm}
\begin{center}
{\bf 2. Summation procedure}
\end{center}

Our summation procedure \cite{0} is based on the fact that
the divergent series
$$
W(g)=\sum\limits_{N=N_0}^{\infty} W_N (-g)^N\,
\eqno(2)
$$
whose coefficients have asymptotic behavior
$W^{as}_N=c\,a^N \Gamma(N+b)$, after the Borel
transformation
$$
W(g)=\int\limits_{0}^{\infty} dx e^{-x} x^{b_0-1} B(gx)\,,\qquad
B(z)=\sum\limits_{N=N_0}^{\infty} B_N (-z)^N\,,\qquad
B_N=\frac{W_N}{\Gamma(N+b_0)}\,
\eqno(3)
$$
(where $b_0$ is an arbitrary parameter) and conformal
mapping\,\footnote{\,This conformal mapping is
different from that used in \cite{3,4}. Its advantage
consists in the more slow growth of random errors in the
coefficients $U_N$ of resummed series (4) and "super-stability"
of the algorithm with respect to smooth errors \cite{0}.
}
$z=u/(1-u)a$ reduces to a convergent power series in $u$
with coefficients
$$
U_0=B_0\,,\qquad U_N=\sum\limits_{K=1}^{N} \frac{B_K}{a^K}
   (-1)^K C_{N-1}^{K-1}\qquad (N\ge 1)\,,
\eqno(4)
$$
whose asymptotics at $N\to \infty$
$$
U_N=U_\infty N^{\alpha-1}\,,
\qquad
U_\infty=\frac{W_\infty}{a^\alpha \Gamma(\alpha) \Gamma(b_0+\alpha)}
\eqno(5)
$$
is related with the strong coupling asypmptotic behavior of
the function $W(g)$,
$$
W(g)=W_\infty g^\alpha\qquad (g\to\infty) \,.
\eqno(6)
$$
The coefficients $U_N$ for $N\alt 40$ are calculated
straightforwardly by Eq.(4) and then are
continued according to power law (5) in order to avoid the catastrophic
increase in errors \cite{0}. Thus, all the coefficients of the
convergent series are known and this series can be summed with
(in principle) arbitrary accuracy. This completely
removes the problem of the
dependence of the results on variation in the summation
procedure, which is the main disadvantage of the commonly
accepted methods.  The interpolation is
performed for the reduced coefficient function
$$
F_N=\frac{W_N}{W_N^{as}}=
    1+\frac{ A_1}{N-\tilde N}+\frac{ A_2}{(N-\tilde N)^2}+\ldots+
    \frac{ A_K}{(N-\tilde N)^K} +\ldots  \,
\eqno(7)
$$
by truncating the series and choosing the coefficients $A_K$
from the correspondence with the known values of the coefficients
$W_{L_0}$, $W_{L_0+1},\ldots$, $W_{L}$. The Lipatov asymptotics
$W_N^{as}$  is taken in the optimal form
 $W_N^{as}=c a^N N^{b-1/2} \Gamma(N+1/2)$
\cite{0}\,\footnote{\,We tried another parametrizations
of the form $W_N^{as}=c a^N N^{\tilde b} \Gamma(N+b-\tilde b)$
but the results were practically the same,
if the same principle
was used for restriction of the set of interpolations.},
 and the parameter $\tilde N$ is used to analyze
uncertainty in the results.  The $L_0$ value sometimes does not
coincide with $N_0$ appearing in Eq.2. Indeed, the coefficient
function $W_N$ continued to the complex plane has a singularity
at the point $N=\alpha$, where $\alpha$ is the exponent of the
strong-coupling asymptotics (6)
\cite{0}. If the exponent
$\alpha$ is larger than $N_0$, the interpolation with the use of
all the coefficients is inapplicable: it is necessary to set
$$
W(g)=W_{N_0}g^{N_0} +\ldots + W_{N_1}g^{N_1} + \tilde W(g)\,,
\qquad N_1=[\alpha]\,,
\eqno(8)
$$
produce summation of the series for $\tilde W(g)$, and add the
contribution from the separated terms; thus, the value $[\alpha]
+ 1$ ($[\ldots]$ is  the integer part of a number) is taken for
 $L_0$.  Analysis of the two-dimensional case \cite{24} shows that
 $\alpha$ is larger than $N_0$ for almost all the functions.

One can see that realization of this programe includes (at
the intermediate stage)
determination of the strong coupling asymptotics
of the function $W(g)$. A large accuracy of this asymptotics is
not necessary for summation in the region $g\sim 1$ and
its rough estimation is sufficient.

Following the tradition, we sum the series not only
for the functions $\beta(g)$,  $\eta(g)$, $\eta_2(g)$, but also
for the functions $\nu^{-1}(g)=2+\eta_2(g)-\eta(g)$
and $\gamma^{-1}(g)=1-\eta_2(g)/(2-\eta(g))$
in order to verify  self-consistency of the
results.
A set of possible interpolations was restricted by two natural
requirements \cite{24}: (a) the interpolation curve  comes
smoothly through the known points and does not have essential
kinks for non-integer $N$; (b)  large $N$ asymptotic behavior
is reached sufficiently quickly, and nonmonotonities
at large $N$ are on the same scale, as a relative difference of
the last known coefficient from the Lipatov asymptotics.

\vspace{6mm}
\begin{center}
{\bf 3. The polymer case ($n=0$)}
\end{center}

Initial information is given by expansions  \cite{2,4}
$$
\beta(g)=-g+g^2-0.4398148149 g^3+ 0.3899226895 g^4-0.4473160967 g^5
+0.63385550 g^6
 $$
 $$ -1.034928 g^7+\ldots+ c a^N \Gamma(N+b) g^N+\ldots \,,
$$
$$
\eta(g)=
(1/108)g^2+0.000 771 3750 g^3+
0.001 589 8706 g^4- 0.000 660 6149 g^5 \qquad
\eqno(9)
$$
$$
+0.001 410 3421 g^6 -0.001 901 867 g^7+\ldots+ c' a^N
\Gamma(N+b') g^N +\ldots\,,
$$
$$
\eta_2(g)=-(1/4) g + (1/16)g^2 - 0.035 767 2729 g^3+
0.034 374 8465 g^4 - 0.040 895 8349 g^5
$$
$$
+0.059 705 0472 g^6 -0.099 284 87 g^7+\ldots+ c'' a^N \Gamma(N+b)
g^N+ \ldots \,,
$$
with the parameters  \cite{5}
$$
a=0.16624600\,,\quad b=b'+1=4\,,
\qquad c=0.085489\,,\quad c'=0.0028836\,,\quad
c''=0.010107\,.
\eqno(10)
$$
Below we discuss some technical details of the summation
procedure.

\vspace{2mm}

{\it Function $\beta(g)$.}  All the interpolations with $L_0=1$
are unsatisfactory: the interpolation curves rapidly achieving
their asymptotic behavior exhibit a sharp kink in the
interval $1<N<2$, indicating a singularity in this
interval. Estimation of the strong-coupling asymptotics
yields $\alpha\approx 1$, confirming
the singularity at $N\approx 1$ and indicating that the choice
$L_0=2$ should be made.
In this case, the interpolation curves with $\tilde N<-0.9$
exhibit significant nonmonotonicity at large $N$, and the curves
with $\tilde N>1.1$ have a kink in the interval $2<N<3$ (see
Fig.1).
\begin{figure}
\centerline{\includegraphics[width=7.0 in]{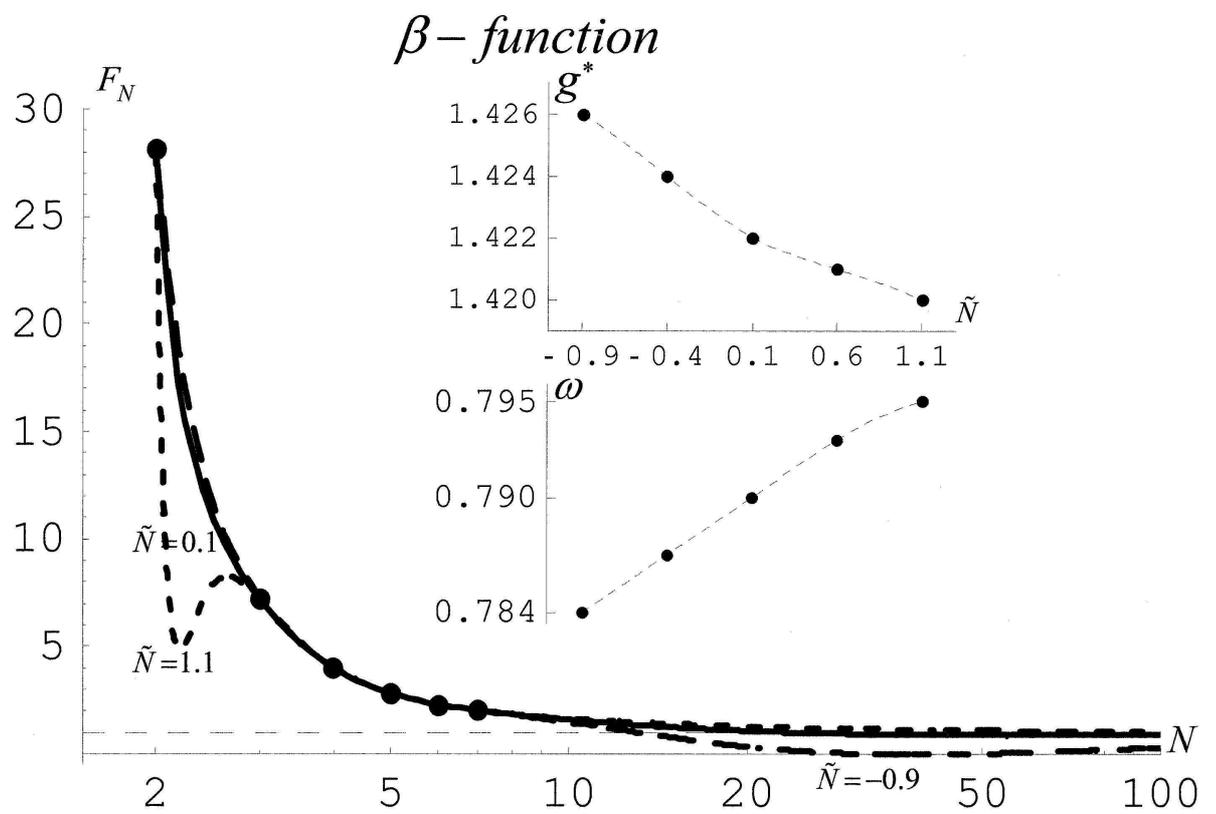}}
\caption{Interpolation curves for the expansion coefficients of
$\beta(g)$  and summation results for $g^*$ and $\omega$. }
\label{fig1}
\end{figure}
Thus, the "natural" interpolations correspond to the
interval $-0.9<\tilde N<1.1$.  The summation results are shown in
the inset in Fig.1, which indicate that
$$
g^*=1.420\div 1.426\,,\qquad \omega=0.784\div 0.795   \,.
\eqno(11)
$$
The $g^*$ value is in agreement with the results of
early works ($g^*=1.421\pm 0.004$ \cite{2},
$g^*=1.421\pm 0.008$ \cite{3}) and in a certain conflict with the
more recent evaluation $g^*=1.413\pm 0.006$ \cite{4}.

\vspace{2mm}

{\it  Function $\eta(g)$. }  According to Eq. (3), the
expansion for $\eta(g)$ begins with $g^2$. We fail to obtain
satisfactory interpolations with $L_0 = 2$: the curves rapidly
approaching large $N$ asymptotic behavior exhibit a kink in the
interval $2<N<3$,  indicating that the exponent $\alpha$
lies in the same interval. Indeed, the estimate of
strong-coupling behavior  gives $\alpha\approx 2$ and suggests
the choice $L_0=3$.
In this case, the satisfactory interpolation curves (see Fig.2,a)
\begin{figure}
\centerline{\includegraphics[width=7.5 in]{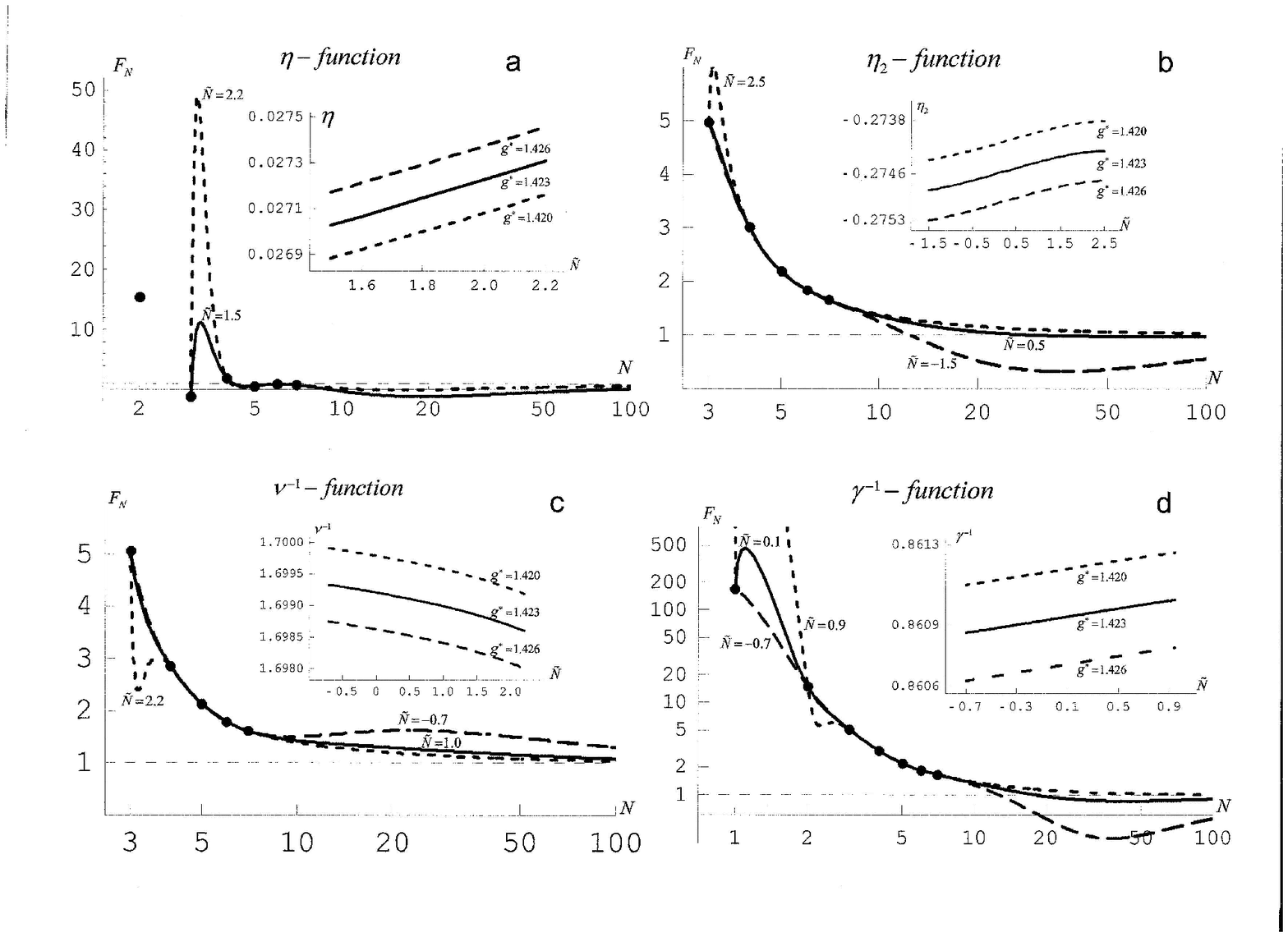}}
\caption{Interpolation curves for the expansion coefficients of
functions $\eta(g)$ (a), $\eta_2(g)$ (b), $\nu^{-1}(g)$ (c) and
 $\gamma^{-1}(g)$ (d).  The insets show the summation results at
  $g = g^*$.}
\label{fig2}
\end{figure}
exist only for $1.5<\tilde N<2.2$. They could be considered
unsatisfactory due to a kink for $3<N<4$; however,
the curves of such a shape provide the exact $\eta$ value in
the two-dimensional case  \cite{24}. In our opinion, such
interpolations are allowable  because the amplitude of
oscillations of the coefficient function is on the order of the
amplitude of oscillations of the known coefficients. The
summation results are shown in the inset in Fig.2,a and give
$\eta=0.0269\div 0.0275$.

 \vspace{2mm}

{\it Functions   $\eta_2(g)$, $\nu^{-1}(g)$ и $\gamma^{-1}(g)$.}
Rough estimates of the strong coupling behavior for
$\eta_2(g)$ and $\nu^{-1}(g)$ gives $\alpha\approx 2$
but in general the results are not self-consistent and violate
the relations between RG functions.
Analysis of the 2D case \cite{24}
shows that it is related with
specific features of $\eta(g)$:  due to small expansion
coefficients, this function is small for $g\alt 10$, but
grows rapidly for large $g$. As a result, asymptotic behavior of
$\nu^{-1}(g)$ and $\eta_2(g)$ contains a mixture of $g$ and
$g^2$ terms, which is difficult to analyze numerically. Therefore,
summation of the series for $\eta_2(g)$ and $\nu^{-1}(g)$ is
performed at $L_0=3$\,\,\footnote{\,Summation at $L_0=2$ gives
practically the same results but with lesser uncertainty.}
 in order to take into account a
possible singularity at $N\approx 2$, while the series for
$\gamma^{-1}(g)$ is summed at $L_0 = 1$, but without the
restriction of kinks for noninteger $N$ (due to
relation $\gamma^{-1}(g)=1+\eta_2(g)/(2- \eta(g))$ its coefficient
function is expected to be regular for $N\ge 1$, but containing a
smeared singularity at $N\approx 1$).  Figures 2,b--d show the
allowable interpolations and summation results. The latter are
presented in Table 1 and compared with results by other authors

 \begin{center}
\hspace{10mm} {\em T a b l e  1.} \\
Critical exponents for the polymer case ($n=0$) from the field
theory
\vspace{2mm}

\begin{tabular}{||c|c|c|c|c|c|c||}
\hline
  &  & &  & & &      \\
{   }        & BNM \cite{2}  & LG--ZJ \cite{3} &G--ZJ \cite{4}&
Kl \cite{9}& J--Kl \cite{10} & Present work \\  &  &  &  &  & &
\\ \hline & & & & & & \\ $\gamma$ & $1.161(3)$ & $1.1615(20)$  &
$1.1596(20)$ & $1.161$ & $1.1604(8) $ & $1.1615(4)$ \\ & &  & &
& & \\ $\nu$ & $ 0.588(1)$ & $ 0.5880(15)$ & $0.5882(11)$ &
$0.5883$ & $0.5881(8) $ & $0.5886(3)$ \\ &  & & & & & \\ $\eta$ &
$ 0.026(14)$  & $0.027(4)$ & $0.0284(25)$  & $0.0311(10)$ & $
0.0285(6) $ & $0.0272(3)$ \\ & &  & &  & & \\ $\eta_2$ & $-0.274
(10)$  & $-0.2745(35)$  & ------ & ------ & ------- &
$-0.2746(7)$ \\ &  & & &  & & \\ $\omega$ & $0.794(6)$ &
$0.800(40)$  & $0.812 (16)$ & $0.810$ & $0.803(3)$ & $0.790(6)$
\\ &  & & &  & &
\\ $g^*$ & $1.421(4)$ & $1.421(8)$ & $1.413(6)$ & ------
& ------ & $1.423(3)$
\\ & &  &   & & & \\
\hline \end{tabular}
\end{center}
\vspace{2mm}

The results for $\nu$ obtained by summation of different series
(in view of relations $\gamma=\nu(2-\eta)$,
$\nu^{-1}=2+\eta_2-\eta$,
$\nu=(1-\gamma)/\eta_2$) are presented in
Table 2. The fourth estimate is rather inaccurate and will be
ignored, while the first three estimates practically coincide;
the relative shift of the central values for them can be
considered as the scale of the non-controllable systematic error,
$$
\delta_{syst} \approx 0.0002 \,,
\eqno(12)
$$
which appears because the "natural" interpolations for different
interdependent functions are not completely consistent. For the
two-dimensional case \cite{24}, this effect is the main source of
the error: a similar estimate gives $\delta_{syst} \approx 0.05$,
which is larger than the natural summmation error for most
functions. At the present case $\delta_{syst}$ is rather small.

  \begin{center}
\hspace{0mm} {\em T a b l e  2.} \\
Different estmates for the exponent $\nu$
\vspace{2mm}
\begin{tabular}{||c|c|c||}
\hline
 & &      \\
Series         & Interval for $\nu$  & Central value \\
  & &  \\
  \hline  & &
\\ $\nu^{-1}(g)$ & $0.5883\div 0.5889$ & $0.5886$
\\ & &
\\  $\gamma^{-1}(g)$, $\eta(g)$ & $\quad 0.5885\div 0.5891\quad$
& $\quad 0.5888\quad$
\\  & &
\\ $\eta_2(g)$, $\eta(g)$ & $ 0.5884\div 0.5892$  & $0.5888$
\\ & &
\\ $\gamma^{-1}(g)$, $\eta_2(g)$ & $0.5848\div 0.5916$ & $0.5882$
\\  & & \\
 \hline
 \end{tabular}
 \end{center}

\vspace{6mm}
\begin{center}
{\bf 4. The Ising universality class ($n=1$)}
\end{center}

Initial information is given by expansions  \cite{2,4}
$$
\beta(g)=-g+g^2-0.422 496 5707 g^3+ 0.351 069 5978 g^4-
0.376 526 8283 g^5 +0.495 547 51 g^6
$$
$$ -0.749 689 g^7+\ldots+ c a^N \Gamma(N+b) g^N+\ldots \,,
$$
$$
\eta(g)=(8/729)g^2
+0.000 914 2223 g^3+ 0.001 796 2229 g^4-
0.000 653 6980 g^5
\qquad \eqno(13)
$$
$$ +0.001 387 8101 g^6
-0.001 697 694 g^7+\ldots+ c' a^N \Gamma(N+b') g^N +\ldots\,,
$$
$$
\eta_2(g)=-(1/3) g + (2/27)g^2 - 0.044 310 2531 g^3+
0.039 519 5688 g^4 - 0.044 400 3474 g^5
$$
$$
+0.060 363 4414 g^6 -0.093 249 48 g^7+\ldots+ c'' a^N \Gamma(N+b)
g^N+ \ldots \,,
$$
with the parameters  \cite{5}
$$
a=0.147 774 22\,,\quad b=b'+1=4.5\,,\quad
c=0.039962\,,\quad c'=0.0017972\,,\quad
c''=0.0062991\,.
\eqno(14)
$$
The situation is qualitatively analogous to the previous case,
and we use the same values for the parameter $L_0$, i.e.
$L_0=1$ for $\gamma^{-1}(g)$, $L_0=2$ for $\beta(g)$,
$L_0=3$ for other functions. Admissible interpolations
correspond to the intervals $-1.0<\tilde N<1.2$  for $\beta(g)$,
$1.6<\tilde N<2.2$  for $\eta(g)$,
$-0.5<\tilde N<2.1$  for $\nu^{-1}(g)$,
$-6.2<\tilde N<2.6$  for $\eta_2(g)$,
$-1.1<\tilde N<0.95$  for $\gamma^{-1}(g)$, and
the appearance of the interpolation curves is visually close to
that for a case $n=0$ (Figs.1,2). The results are presented in
Table 3.

  \begin{center}
\hspace{10mm} {\em T a b l e  3.} \\
Critical exponents for the Ising case ($n=1$) from the field
theory

\begin{tabular}{||c|c|c|c|c|c|c||}
\hline
  &  & &  & & &      \\
{   }        & BNM \cite{2}  & LG--ZJ \cite{3} &G--ZJ \cite{4}&
Kl \cite{9}& J--Kl \cite{10} & Present work \\
&  &  &  &  & &  \\
\hline & & & & & &
\\ $\gamma$ & $1.241(4)$ & $1.2405(15)$  & $1.2396(13)$ &
$1.241$ & $1.2403(8) $ & $1.2411(6)$
\\ & &  & &  & &
\\ $\nu$ & $ 0.630(2)$ & $ 0.6300(15)$ &
$0.6304(13)$ & $0.6305$ & $0.6303(8) $ & $0.6306(5)$
\\ &  & & & & &
\\ $\eta$ & $ 0.031(11)$  & $0.032(3)$ & $0.0335(25)$  &
$0.0347(10)$ & $ 0.0335(6) $ & $0.0318(3)$
\\ & &  & &  & & \\
$\eta_2$ & $-0.382 (5)$  & $-0.3825(30)$  & ----- &
------ & ------ & $-0.3832(8)$
\\ &  & & &  & &
\\ $\omega$ & $0.788(3)$ & $0.790(30)$  & $0.799 (11)$ &
$0.805$ & $ 0.792(3) $ & $0.782(5)$
\\ &  & & &  & &
\\ $g^*$ & $1.4160(15)$ & $1.416(5)$ & $1.411(4)$ & ------
& ------ & $1.4185(25)$
\\ & &  &   & & & \\
\hline \end{tabular}
\end{center}

\vspace{6mm}
\begin{center}
{\bf 5. The XY universality class ($n=2$)}
\end{center}

The detailed discussion of this case is given in the paper
\cite{25}. For completeness, we present here the final
results of this study (Table 4).

  \begin{center}
\hspace{10mm} {\em T a b l e  4.} \\
Critical exponents for the XY (or helium) case ($n=2$) from the
field theory

\begin{tabular}{||c|c|c|c|c|c|c||}
\hline
  &  & &  & & &      \\
{   }        & BNM \cite{2}  & LG--ZJ \cite{3} &G--ZJ \cite{4}&
Kl \cite{9}& J--Kl \cite{10} & Present work
\\  &  &  &  &  & &  \\
\hline & & & & & &
\\ $\gamma$ & $1.316(9)$ & $1.3160(25)$  & $1.3169(20)$ &
$1.318$ & $1.3164(8) $ & $1.3172(8)$
\\ & &  & &  & &
\\ $\nu$ & $ 0.669(3)$ & $ 0.6695(20)$ &
$0.6703(15)$ & $0.6710$ & $0.6704(7) $ & $0.6700(6)$
\\ &  & & & & &
\\ $\eta$ & $ 0.032(15)$  & $0.033(4)$ & $0.0354(25)$  &
$0.0356(10)$ & $ 0.0349(8) $ & $0.0334(2)$
\\ & &  & &  & & \\
$\eta_2$ & $-0.474 (8)$  & $-0.4740(25)$  & ------ &
------ & ------- & $-0.4746(9)$
\\ &  & & &  & &
\\ $\omega$ & $0.780(10)$ & $0.780(25)$  & $0.789 (11)$ &
$0.800$ & $0.784(3)$ & $0.778(4)$
\\ &  & & &  & &
\\ $g^*$ & $1.406(5)$ & $1.406(4)$ & $1.403(3)$ & ------
& ------ & $1.408(2)$
\\ & &  &   & & & \\
\hline \end{tabular}
\end{center}

\vspace{6mm}
\begin{center}
{\bf 6. The Heisenberg universality class ($n=3$)}
\end{center}

Initial information is given by expansions  \cite{2,4}
$$
\beta(g)=-g+g^2-0.383 226 2015 g^3+ 0.282 946 6813 g^4
-0.270 333 30 g^5 +0.312 5559 g^6
$$
$$
-0.414 861 g^7+\ldots+
c a^N \Gamma(N+b) g^N+\ldots \,,
$$
$$
\eta(g)=(40/3267)g^2
+ 0.001 020 0000 g^3+ 0.001 791 9257 g^4- 0.000 504 0977 g^5
\qquad \eqno(15)
$$
$$ +0.001 088 3237 g^6 -0.001 111499 g^7+\ldots+
c' a^N \Gamma(N+b') g^N +\ldots\,,
$$
$$
\eta_2(g)=-(5/11) g + (10/121)g^2 - 0.052 551 9564 g^3+
0.039 964 0005 g^4 - 0.041 321 9917 g^5
$$
$$
+0.049 092 9344 g^6 -0.067 086 30 g^7+\ldots+ c'' a^N \Gamma(N+b)
g^N+ \ldots \,,
$$
with the parameters  \cite{5}
$$
a=0.12090618\,,\quad b=b'+1=5.5\,,\quad
c=0.0059609\,,\quad c'=0.0003656\,,\quad
c''=0.0012813\,.
\eqno(16)
$$
The same values for $L_0$, as in previous
cases, were used. Admissible interpolations
correspond to the intervals $-1.0<\tilde N<1.6$  for $\beta(g)$,
$1.6<\tilde N<2.3$  for $\eta(g)$
$0.4<\tilde N<2.0$  for $\nu^{-1}(g)$,
$-0.6<\tilde N<2.2$  for $\eta_2(g)$,
$0.5<\tilde N<0.95$  for $\gamma^{-1}(g)$.
The results are presented in Table 5.

  \begin{center}
\hspace{10mm} {\em T a b l e  5.} \\
Critical exponents for the Heisenberg case ($n=3$) from the field
theory
\vspace{2mm}

\begin{tabular}{||c|c|c|c|c|c|c||}
\hline
  &  & &  & & &      \\
{   }        & BNM \cite{2}  & LG--ZJ \cite{3} &G--ZJ \cite{4}&
Kl \cite{9}& J--Kl
\cite{10} & Present work \\  &  &  &  &  & &  \\
\hline & & & & & &
\\ $\gamma$ & $1.390(10)$ & $1.386(4)$  & $1.3895(50)$ &
$1.390$ & $1.3882(10) $ & $1.3876(9)$
\\ & &  & &  & &
\\ $\nu$ & $ 0.705(5)$ & $ 0.705(3)$ &
$0.7073(35)$ & $0.7075$ & $0.7062(7) $ & $0.7060(7)$
\\ &  & & & & &
\\ $\eta$ & $ 0.031(22)$  & $0.033(4)$ & $0.0355(25)$  &
$0.0350(10)$ & $ 0.0350(8) $ & $0.0333(3)$
\\ & &  & &  & & \\
$\eta_2$ & $-0.550 (12)$  & $-0.5490(35)$  & ------ &
------ & ------- & $-0.5507(12)$
\\ &  & & &  & &
\\ $\omega$ & $0.780(20)$ & $0.780(20)$  & $0.782 (13)$ &
$0.797$ & $0.783(3)$ & $0.778(4)$
\\ &  & & &  & &
\\ $g^*$ & $1.392(9)$ & $1.391(4)$ & $1.390(4)$ & ------
& ------ & $1.393(2)$
\\ & &  &   & & & \\
\hline \end{tabular}
\end{center}
\vspace{2mm}

\vspace{6mm}
\begin{center}
{\bf 7. Discussion}
\end{center}

One can see from Tables 1,3,4,5 that there is a
good correspondence between the different field theory
estimations.  A surprisingly good agreement takes place between
our results and estimation by Le Guillou -- Zinn-Justin \cite{3}:
the typical difference of the central values is less than 0.0010,
in spite of rather conservative estimation of errors given in
\cite{3}.  This coincidence is not in any degree incidental: the
authors of \cite{3} carried out interpolation of the coefficient
function in order to predict one or two of  unknown
expansion coefficients and used them to give some kind of the
expert prediction, but were induced to allow rather
large uncertainty of results due to their strong dependence
on variation of the summation procedure.
On the other hand, recent reevaluation in
\cite{4} looks somewhat artificial and has a tendency to shift
the results beyond their natural range;
in particular, the shift of
$g^*$ in comparison with \cite{3} is always made in the direction
opposite to ours (Tables 1,3,4,5)\,\footnote{\,It should be noted
that \cite{3,4} and the present paper use the same information
for $\beta(g)$.}. A good agreement can be seen also with
variational perturbation theory  \cite{9}; it is especially
pleasant that taking into account the large order perturbation
behavior and more elaborated estimation of errors \cite{10} makes
the results  more close to
ours. A small disagreement still remains for the exponent $\eta$
but it is on the same level as violation of the relation
$\gamma=\nu(2-\eta)$ for the central values of \cite{10}.

Now let us discuss the correspondence of our results with other
information on the critical exponents, provided by physical
experiment, Monte Carlo simulations (MC) and high temperature
series (HT) \cite{20}.

{\it Case $n=3$}. Overall scattering of the MC and HT results is
rather large and in this extent they agree with Table 5.
There is a tendency to a small disagreement between our and
the most
recent MC results ($\gamma=1.3960(10)$, $\nu=0.7112(5)$
\cite{234}) but the latter are in the same disagreement with
the physical experiments, whose results for $\gamma$ are grouped
around value $1.386$ (see Tables 24,25 in \cite{20}).
Analogously, the experimental results for the exponent $\beta$
suggest the mean value $0.365$ \cite{20}
in the good agreement with our estimate $\beta=0.3648(4)$
(following from Table 5) and in
the worse agreement
with  value $\beta=0.3689(3)$ given in \cite{234}.

{\it Case $n=2$}. A situation is analogous to the previous
case. Overall scattering of the MC and HT results is
rather large (see Fig.1 in \cite{25}) but the recent results have
a tendency to contradict Table 4 ($\gamma=1.3178(2)$,
$\nu=0.6717(1)$ $\eta=0.0381(2)$ \cite{17}). Simultaneously they
contradict the experiments in liquid helium, i.e. value
$$
\nu=0.6705\pm 0.0006\,,
\eqno(17)
$$
obtained by the measurements of superfluid density from the
second sound velocity \cite{31}, and the results
$$
\alpha=-0.01285\pm 0.00038 \,,\qquad \nu=0.67095(13)\quad
\cite{32}\,,
$$
$$
\alpha=-0.01056\pm 0.00038 \,,\qquad \nu=0.6702(1)
\quad\cite{33} \,,
\eqno(18)
$$
$$
\alpha=-0.0127\pm 0.0003\,, \qquad \nu=0.6709(1) \quad
\cite{34}
$$
obtained in the satellite measurements of the thermal capacity
(the relation $\alpha=2-d\,\nu$ was used).

{\it Case $n=1$}. In this case, the  HT and MC results are
numerious (see Tables 3,5 in \cite{20}) and
can be summarize as
$$
\gamma=1.2372(5)
$$
$$
\nu=0.6301(4)
\eqno(19)
$$
$$
\eta=0.0364(5)
$$
(see Eq.3.2 in \cite{20}).
One can see from the Table 3 that beautiful consensus was reached
for the exponent $\nu$; on the other hand, values for $\gamma$
and $\eta$ in (19) are in the meaningful contradiction with Table
3. The experimental results have large uncertainty and cannot
compete with theoretical predictions.

{\it Case $n=0$}. In this case, precise results for the exponent
$\nu$ can be obtained by direct study of self-avoiding walks
on the lattice; due to simplicity of the algorithm, a good
statistics can be gathered. The most recent results
($\nu=0.5876(2)$ \cite{100}, $\nu=0.5874(2)$ \cite{925},
$\nu=0.58758(7)$ \cite{125}) appear in a slight contradiction
with our result in Table 1. This contradiction is not very
significant and we can avoid it by allowing more wide set of
interpolations and extending the error bars;
however, somewhat "unnatural" interpolation curves
should be used for it. The results for the exponent
$\gamma$ are essentially less precise \cite{20} and cannot
compete with Table 1.

We can conclude that the general situation is satisfactory
but there is a cause for anxiety related with the exponents
$\gamma$ and $\eta$ in the Ising case. Disagreement on the scale
0.003 is essentially larger than uncertainty of the recent
field theoretical estimations (Table 3) and that of Monte Carlo
results. It is also essential that the latter are
obtained by different researches and not related
with a specific group.
At present, the origin of this disagreement is unclear
and further investigations are necessary.

\vspace{3mm}

This work is partially supported by RFBR  (grant 06-02-17541).

\end{document}